\documentstyle[prl,aps,epsf,twocolumn,floats,color]{revtex}

\begin{document}

\draft

\wideabs{

\title{
Ordering Process and Its Hole Concentration Dependence of \\
the Stripe Order in La$_{2-x}$Sr$_{x}$NiO$_{4}$
}

\author{
R. Kajimoto,$^{1}$ T. Kakeshita,$^{2,}$\cite{kake_ad}
H. Yoshizawa,$^{2}$ T. Tanabe,$^{3}$ T. Katsufuji,$^{3,}$\cite{katsu_ad}
and Y. Tokura$^{3}$
}

\address{
$^{1}$Department of Physics, Ochanomizu University, Bunkyo-ku,
Tokyo 112-8610, Japan
$^{2}$Neutron Scattering Laboratory, Institute for Solid State
Physics, University of Tokyo, Tokai, Ibaraki 319-1106 \\
$^{3}$Department of Applied Physics, University of Tokyo, Bunkyo-ku,
Tokyo 113-8656
}

\date{
\today
}

\maketitle

\begin{abstract}

Ordering process of stripe order in La$_{2-x}$Sr$_{x}$NiO$_{4}$ with $x$
being around 1/3 was investigated by neutron diffraction
experiments. When the stripe order is formed at high temperature,
incommensurability $\epsilon$ of the stripe order has a tendency to show
the value close to 1/3 for the samples with $x$ at both sides of
1/3. With decreasing temperature, however, $\epsilon$ becomes close to
the value determined by the linear relation of $\epsilon = n_{\rm h}$,
where $n_{\rm h}$ is a hole concentration. This variation of the
$\epsilon$ strongly affects the character of the stripe order through
the change of the carrier densities in stripes and antiferromagnetic
domains.

\end{abstract}

\pacs{
PACS numbers: 75.25.+z, 71.27.+a, 71.45.Lr, 74.80.Dm}

}

%\section{Introduction}

A unique feature of stripe order in high-$T_{\rm c}$
cuprates~\cite{stripe_nature} has been drawing considerable attention of
a number of experimentalists as well as theorists. A similar stripe
order observed in a hole-doped nickelate
La$_{2-x}$Sr$_{x}$NiO$_{4+\delta}$ makes this compound a good candidate
for a study of the stripe order because its isomorphic crystal structure
with high-$T_{\rm c}$ cuprates, and further stability of the stripe
order compared with cuprates. With decreasing temperature, the charge
stripe is formed first at $T = T_{\rm CO}$, and then the spins order
antiferromagnetically with antiphase domains at the charge stripes at
the lower temperature $T_{\rm
N}$~\cite{tranquada95,wochner98,sachan95,tranquada96,yoshizawa00}. The
modulation vector of the spin order is given by $\bbox{g}_{\rm spin} =
\bbox{Q}_{\rm AF} \pm (\epsilon,0,0)$, where $\bbox{Q}_{\rm AF} =
(1,0,0)$ is the wave vector for a simple antiferromagnetic order, while
that of the charge order is $\bbox{g}_{\rm charge} = (2\epsilon,0,0)$ in
the orthorhombic cell.

Although the stripe order in nickelates has been well studied for the
low doped region for $n_{\rm h} \le 1/3$ where $n_{\rm h}$ is a hole
concentration~\cite{tranquada95,wochner98,sachan95,tranquada96,lee97,lee01},
few work has been reported so far for the highly doped region with
$n_{\rm h} > 1/3$. Very recently we have extended the study of the
stripe order with neutron diffraction technique toward the higher doping
region up to $n_{\rm h} = 1/2$~\cite{yoshizawa00}. Surprisingly, we have
observed that the stripe order persists up to $n_{\rm h} = 1/2$, and the
incommensurability $\epsilon$ is approximately linear against $n_{\rm
h}$. From the $n_{\rm h}$ dependence of the onset temperatures of the
charge stripe and spin order, $T_{\rm CO}$, $T_{\rm N}$, and the
correlation length of the stripe order, we argued that the stripe is
most stable at $n_{\rm h} =1/3$. In order to further elucidate the
nature of the stripe order in nickelates, especially, its ordering
process and the influence of the commensurability at $n_{\rm h} =1/3$,
we have performed a detailed neutron diffraction study on the three
Sr-doped nickelate samples La$_{2-x}$Sr$_{x}$NiO$_{4}$ with $x < 1/3$,
$x=1/3$, and $x > 1/3$. 

%This is the first systematic study on the ordering process of the stripe
%order and its hole concentration dependence keeping the importance of
%the magic number $n_{\rm h}=1/3$ in mind.

The central results reported in this paper is that three distinct
temperature regions are identified in the ordering process of the charge
stripe order. The variation of the incommensurability of the charge
stripe order plays a crucial role to determine the character of the
stripes by the variation of the carrier density in the charge
stripes. The character of the stripes also depends on the hole
concentration, and it shows symmetrical behavior around $n_{\rm h} =
1/3$ due to the strong influence of the commensurability at this
concentration. The preliminary results have been reported
elsewhere~\cite{kaji_asr}.

%\section{Experimental procedure}

Single crystal samples studied in the present study were grown by the
floating zone method. The crystal structure is pseudo tetragonal. The
oxygen off-stoichiometry $\delta$ as well as the hole concentration
$n_{\rm h} =x + 2\delta$ were characterized in detail as previously
reported~\cite{katsufuji99}. The calibrated hole concentration $n_{\rm
h}$ for the present samples are 0.289, 0.339, and 0.39. We denote the
samples by $n_{\rm h}$ throughout this report.

The neutron diffraction experiments were performed using triple axis
spectrometer GPTAS installed at the JRR-3M reactor in JAERI, Tokai,
Japan with a fixed incident neutron momentum of 2.57 {\AA}$^{-1}$.  We
chose a combination of horizontal collimators of
40$^{\prime}$-80$^{\prime}$-40$^{\prime}$-80$^{\prime}$ (from
monochromator to detector) for most of scans except for profile scans
for $n_{\rm h} =0.39$ which were measured with
20$^{\prime}$-40$^{\prime}$-40$^{\prime}$-40$^{\prime}$. The crystals
were mounted in Al cans filled with He gas, and were attached to a cold
head of a closed-cycle He gas refrigerator. The temperature was
controlled within an accuracy of 0.2 K. We employ the orthorhombic
setting for convenience of easier comparison with preceding works. All
the measurements were performed on the $(h,0,l)$ scattering plane.

%\section{Results}

%\subsection{Fig.\ \ref{profile}: Temperature dependence of the profiles
%of the charge and spin order peaks}

\begin{figure}
 \centering \leavevmode
 \epsfxsize=0.9\hsize
 \epsfbox{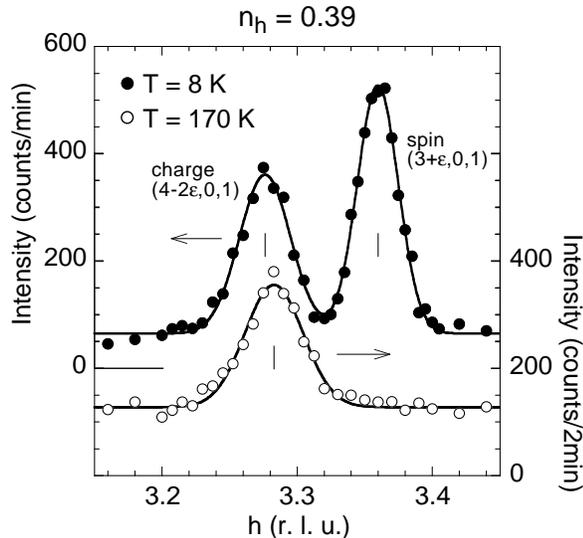}
% \vspace{-2mm}
 \caption{
 Profiles for the charge and spin superlattice peaks observed along
 $(h,0,1)$ for $n_{\rm h} = 0.39$ measured at $T = 10$ K (open symbols)
 and 170 K (closed symbols). Solid lines are the fits to Gaussian and
 vertical bars indicate the peak positions.
 }
 \label{profile}
\end{figure}

Let us begin with the profiles of charge/spin stripe order.  In order to
characterize the stripe order in the $ab$ plane, we performed scans
along the [100] direction. Figure~\ref{profile} shows typical profiles
of the charge and spin superlattice peaks measured along the $(h,0,1)$
line for $n_{\rm h} = 0.39$, which has a larger hole concentration than
$n_{\rm h} =1/3$. At 170 K, only the charge order peak is visible,
reflecting that the charge ordering temperature $T_{\rm CO}$ is higher
than the spin ordering temperature $T_{\rm N}$ for a sample with $n_{\rm
h} > 1/3$ similar to those with $n_{\rm h} < 1/3$. At somewhat lower
$T$, a peak of the spin order appears, and its intensity grows with
decreasing $T$.  At 8 K well-defined charge and spin order peaks are
observed at $(4-2\epsilon,0,1)$ and $(3+\epsilon,0,1)$ with $\epsilon
\simeq 0.36$, respectively. Furthermore, a careful inspection of the
profiles of the charge order peak reveals that the incommensurability
$\epsilon$ has a small $T$ dependence.

%\subsection{Fig.\ \ref{manyTdep}: Temperature dependence of intensity,
%$\epsilon$ and peak width}

%\subsubsection{overview of Fig.\ \ref{manyTdep}.}

\begin{figure*}[t]
 \centering \leavevmode
 \epsfxsize=0.9\hsize
 \epsfbox{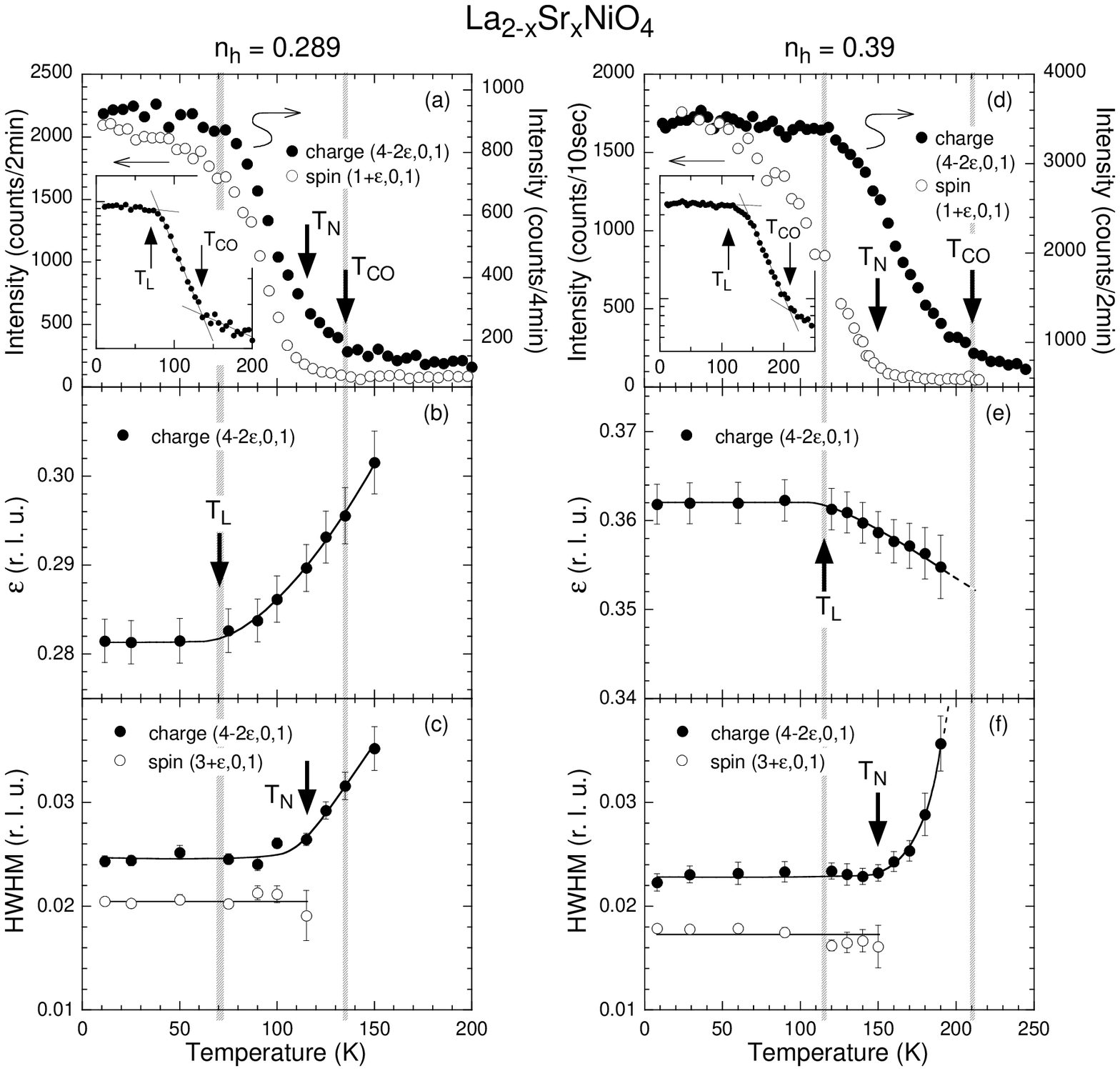}
 \vspace{0mm}
 \caption{
 Temperature dependence of the scattering intensity ((a), (d)), the
 incommensurability $\epsilon$ ((b), (e)), and the peak width (HWHM)
 ((c), (f)) of the charge order peak (closed symbols) and the spin order
 peak (open symbols) for $n_{\rm h} = 0.289$ ((a)--(c)) and $n_{\rm h} =
 0.39$ ((d)--(f)). The intensities of the charge order peaks are
 replotted in logarithmic scale as insets in (a) and (d).
 }
 \label{manyTdep}
\end{figure*}

To examine the ordering process of the stripe order in more detail, we
have carried out a systematic study of the $T$ dependence of the charge
and spin stripe superlattice peaks in samples which have the hole
concentration $n_{\rm h} < 1/3$ and $n_{\rm h} > 1/3$. The $T$
dependences of intensity, $\epsilon$, and width for the two selected
samples with $n_{\rm h} = 0.289$ and 0.39 are summarized in
Fig.~\ref{manyTdep}.

Figures~\ref{manyTdep}(a) and \ref{manyTdep}(d) show the $T$ dependences
of the intensities of the $(4-2\epsilon,0,1)$ charge order peak and
those of the $(1+\epsilon,0,1)$ spin order peak. As mentioned above,
$T_{\rm CO}$ is higher than $T_{\rm N}$ for both
samples~\cite{yoshizawa00}. The $T$ dependence of the intensity exhibits
a distinct anomaly at $T_{\rm L}$, below which the intensity of the
charge order peak saturates. As shown in Figs.~\ref{manyTdep}(b) and
\ref{manyTdep}(e), the incommensurability $\epsilon$ continuously varies
through the spin ordering temperature $T_{\rm N}$, but locks in at
$T_{\rm L}$.  Notice that the lock-in temperature of $\epsilon$ is well
correlated with the saturation temperature of the intensity of the
charge order peak, $T_{\rm L}$.  As for the correlation lengths of the
charge and spin orders, the width of the charge stripe peak continues to
decrease below $T_{\rm CO}$, but saturates around $T_{\rm N}$ as shown
in Figs.~\ref{manyTdep}(c) and \ref{manyTdep}(f), indicating that, once
spin domains between charge stripes establish an antiferromagnetic (AF)
spin order below $T_{\rm N}$, the correlation length of the charge
stripe order ceases to grow. In other words, the charge stripe order is
short-ranged for $T_{\rm N} < T < T_{\rm CO}$, but
forms quasi-long range order below $T_{\rm N}$.  This
fact strongly indicates AF spin correlations are essential to stabilize
charge/spin stripe order, and implies they are also
responsible for the anomalous behaviors of $\epsilon$ and the peak
intensity for $T_{\rm L} < T < T_{\rm CO}$.

%As indicated by vertical dashed lines in Fig.~\ref{manyTdep}, one can
%identify from these results three $T$ regions for the ordering process
%of the charge/spin stripe order below $T_{\rm CO}$: (i) a short-range
%charge stripe order for $T_{\rm N} < T < T_{\rm CO}$, (ii) a quasi-long
%range charge/spin stripe order with a continuous shift of $\epsilon$ for
%$T_{\rm L} < T < T_{\rm N}$, and (iii) a frozen quasi-long range
%charge/spin stripe order with fractional $\epsilon$ below $T_{\rm L}$.

%\subsection{Fig.\ \ref{epsilon}: temperature dependence of
% $\epsilon$ for charge order peaks for all three samples}

%\subsubsection{high $T$ behavior of $\epsilon$: the commensurability
%effect}

\begin{figure}[t]
 \centering \leavevmode
 \epsfxsize=0.9\hsize
 \epsfbox{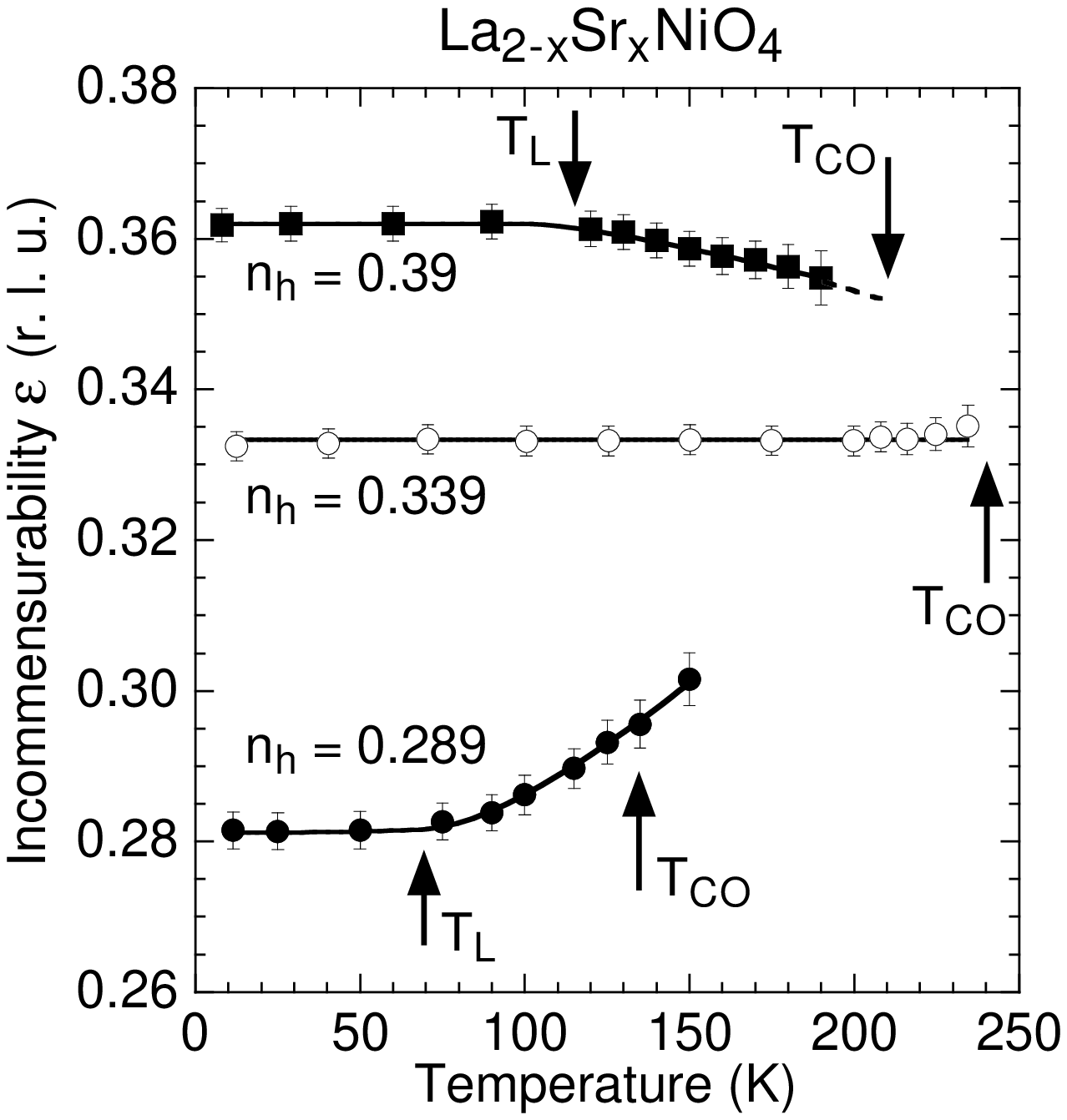}
 \vspace{1mm}
 \caption{
 Temperature dependence of the incommensurability $\epsilon$
 for $n_{\rm h} = 0.289$ (closed circles), $n_{\rm h} = 0.339$ (open
 circles), and $n_{\rm h} = 0.39$ (closed squares).
 }
 \label{epsilon}
\end{figure}

A closer examination of behavior of $\epsilon$ will provide rich
information on the physics of charge/spin stripe order with $n_{\rm h}$
around 1/3.  To visualize the behavior of $\epsilon$ of the {\it
``off-concentration''} samples with $n_{\rm h} = 0.289$ and $n_{\rm h} =
0.39$ against the {\it ``commensurate''} value of $n_{\rm h} = 1/3$,
their $T$ dependences are replotted together with that of the {\it
``commensurate''} $n_{\rm h} = 0.339 \simeq 1/3$ sample in
Fig.~\ref{epsilon}.  The {\it commensurate} $n_{\rm h} =1/3$ sample is
unique because $\epsilon$ exhibits practically no temperature dependence
and stays at $\epsilon = 1/3\ (= n_{\rm h}) $ for all $T$. By contrast,
one can clearly see that the behavior of $\epsilon$ for two {\it
off-concentration} samples is symmetric around $\epsilon = 1/3$.

In our previous study, we established that although the low $T$ value of
$\epsilon$ follows a linear law i.e. $\epsilon = n_{\rm h}$, there
was a small systematic deviation of $\epsilon$~\cite{yoshizawa00}.  From
Fig.~\ref{epsilon}, we find that such a systematic deviation in the {\it
off-concentration} samples is further enhanced at high $T$ near $T_{\rm
CO}$: In either case, $\epsilon$ exhibits a closer value to $1/3$. These
deviations from the $\epsilon = n_{\rm h}$ law at high $T$ for {\it
off-concentration} samples indicate that the charge stripe itself
prefers $\epsilon=1/3$ by tuning the carrier density within the stripes.
We would like to emphasize here that the self-tuning behavior of
$\epsilon$ toward 1/3 at high $T$ in the present {\it Sr-doped} samples
should not be confused with a similar lock-in behavior of $\epsilon$ in
the oxygen-doped samples~\cite{wochner98}. In the excess-oxygen samples,
it is driven by the ordering of the interstitial oxygen atoms and
associated buckling of NiO$_6$ octahedra, whereas the amount of the
excess oxygen is negligible in the present {\it Sr-doped}
samples. Thereby, the tendency of charge stripes favoring $\epsilon
\simeq1/3$ at high $T$ is intrinsic to the stripe ordering in the
Sr-doped system, and we tentatively call it as a {\it commensurability
effect}.

%\section{Discussions}

%\subsection{$n_{\rm st}$ and $\epsilon$}

%\subsubsection{Introduction of $n_{\rm st}$ and its definition}

%\subsubsection{influence of the change of $n_{\rm st}$ on the charge
% stripe}

Next, we discuss the $T$ dependence of $\epsilon$.  With decreasing $T$,
$\epsilon$ decreases for $n_{\rm h} = 0.289$ whereas it increases for
$n_{\rm h} = 0.39$, and is locked below $T_{\rm L}$ for both samples.
Namely, $\epsilon$ becomes closer to $n_{\rm h}$ at low $T$. To describe
the effect of the change of $\epsilon$, we shall introduce the nominal
hole density within charge stripes $n_{\rm st}$. From the observed value
of $\epsilon$, one can evaluate $n_{\rm st}$ by $n_{\rm st} \equiv
n_{\rm h}/\epsilon$.  In terms of $n_{\rm st}$, the observed behavior
of $\epsilon$ indicates that $n_{\rm st}$ strongly deviates from unity
at $T \sim T_{\rm CO}$ due to the commensurability effect at $\epsilon =
1/3$, but it shows a tendency to approach unity upon decreasing $T$.

%When the behavior of the carrier density $n_{\rm st}$ within charge
%stripes is considered, the physics of the $T$ dependence of $\epsilon$
%becomes clearer.  From the observed value of $\epsilon$, one can
%evaluate the carrier density $n_{\rm st}$ within charge stripes by
%$n_{\rm st} \equiv n_{\rm h}/\epsilon$.

When $n_{\rm st} = 1$, all the doped holes are accommodated within the
charge stripes and the stripes become half-filled Mott insulators, and
there are no excess carriers in the system~\cite{katsufuji99}. To the
contrary, when $n_{\rm st} \ne 1$, there are excess electrons ($n_{\rm
st} < 1$) or holes ($n_{\rm st} > 1$) in the system to form the
half-filled stripes. In this situation, there are two possibilities
concerning the location of the excess carriers: the one is they enter
the charge stripe, and the other is they are distributed in a 
NiO$_{2}$ matrix separated by the stripes. Considering the fact that the stripes
are in a short-ranged glassy state at $T \sim T_{\rm CO}$ where $n_{\rm
st}$ strongly deviates from unity, it would be natural to assume that
the excess carriers enter both NiO$_{2}$ matrix and stripes, because it is difficult to
distinguish between the two in the high $T$ glassy-stripe state. To simplify the discussion,
we shall describe such a situation by $n_{\rm st} \ne 1$ in this paper,
although in the strict sense of the definition this quantity is only relevant to the stripes.

The $T$ dependent variation of $n_{\rm st}$ is intimately
related to the development of the AF spin correlations.  It was reported that the AF spin
correlations develop dynamically in the NiO$_{2}$ matrix imediately
below $T_{\rm CO}$, though $T_{\rm
N}$ is well below $T_{\rm CO}$ ~\cite{sachan95,yamamoto98}.  When $n_{\rm st} \ne 1$,
carriers distributed in the NiO$_{2}$ matrix may cause local disturbances
of the AF exchange interactions in the AF spin domains. On the other hand, when $n_{\rm st} =
1$, there is no disturbance to the AF spin correlation. From this
consideration, we suggest that the observed $T$ dependence of $n_{\rm
st}$ is driven by the development of the AF spin correlations at low
$T$'s.  By confining
doped-holes within charge stripes and by adjusting the distances between
the stripes, the AF spin correlations gain the exchange energy, and concomitantly favor
 the half-filled stripe with $n_{\rm st}= 1$.

%\subsubsection{influence of the change of $n_{\rm st}$ on the AF
% domains --- atomic modulations}

The variation of $n_{\rm st}$ driven by AF spin correlations has strong
influence to the lattice distortions caused by the stripe order.  The
extra-electrons or holes supplied to the AF spin domains must cause
local atomic displacements, and disturb the lattice distortions induced
by the charge stripe order, thereby reduce the intensity of the charge
order peak, because it is proportional to square of the coherent
component of the amplitude of the displacements.  Consequently, the
behavior of $\epsilon$ shows a strong correlation with the intensity of
charge stripe peaks, as shown in Figs.~\ref{manyTdep}(a), (b), (d), and
(e).  In particular, the continuous shift of $\epsilon$ gives rise to a
Debye-Waller-like $T$ dependence of the intensity of the stripe order,
i.e. $I \sim e^{-2T/T_{0}}$ for $T_{\rm L} < T < T_{\rm CO}$, as
depicted in the insets of Figs.~\ref{manyTdep}(a) and \ref{manyTdep}(d).
On the analogy of the Debye-Waller factor, one can interpret that such
$T$ dependence indicates the existence of strong fluctuations in the
charge stripe order, being consistent with the continuous change of
$n_{\rm st}$ as well as the correlation length of the stripe order.
  A similar behavior is also observed
in the charge order peak as well as the field induced magnetic order
peak in La$_{2}$NiO$_{4+\delta}$ with $\delta
=2/15$~\cite{wochner98,tranquada97}.

%\subsubsection{influence of the change of $n_{\rm st}$ on the AF
% domains --- spin correlations}

The variation of $n_{\rm st}$ also affects the transport property in the
nickelate system. When $n_{\rm st} \ne 1$, it indicates that charge
stripes and AFM spin domains become doped insulators. In our previous
study~ \cite{yoshizawa00}, we found that $n_{\rm st} < 1$ for $n_{\rm h}
< 1/3$ whereas $n_{\rm st} > 1$ for $n_{\rm h} > 1/3$.  This means that
the character of carriers changes from electron-like to hole-like when
$n_{\rm h}$ crosses $n_{\rm h} = 1/3$, as was evidenced by the previous
Hall coefficient study~\cite{katsufuji99}. The doped carriers give rise
to a relatively high conductivity in the high $T$ region. As $n_{\rm
st}$ gradually becomes unity with lowering $T$, however, the half-filled
stripes segregate within the AF spin-ordered NiO$_{2}$ matrix, and the
 conductivity is continuously reduced. This picture well
explains the fact that the resistivity of the nickelate system does not
exhibit a steep upturn at $T_{\rm CO}$, but it rather shows crossover
behavior except for the $n_{\rm h} = 1/3$
sample~\cite{katsufuji99,cheong94}.

%\subsubsection{competition between the commensurability effect and AF
%spin correlations}

From the results observed in the present study and discussions presented
above, we conclude that the distance of stripes and the
incommensurability $\epsilon$ are determined by two competing effects:
the commensurability effect which favors $\epsilon = 1/3$, and the AF
spin correlation which favors $\epsilon = n_{\rm h}$. The former is
important just below $T_{\rm CO}$ while the latter becomes dominant as
the AF spin correlation develops.  This competition leads to three
distinct $T$ regions which can be identified in the ordering process of
charge/spin stripe order as indicated by vertical dashed lines in
Fig.~\ref{manyTdep}: (i) a disordered stripe with no static order for $T
> T_{\rm CO}$, (ii) a fluctuating charge/spin stripe order with a
continuous shift of $\epsilon$ for $T_{\rm L} < T < T_{\rm CO}$, and
(iii) a frozen quasi-long range charge/spin stripe order with fractional
$\epsilon$ below $T_{\rm L}$.  Right below $T_{\rm CO}$, $\epsilon$
shows a tendency having a value close to 1/3 for the samples with $x$ at
both sides of 1/3 due to the commensurability effect for $\epsilon
=1/3$.  This effect takes place at the expense of a deviation of the
stripe carrier density $n_{\rm st}$ from unity.  With decreasing $T$,
however, AF spin correlations favor half-filled stripes to gain the
exchange energy.  Consequently, $\epsilon$ becomes close to the value
determined by the linear relation of $\epsilon = n_{\rm h}$, and $n_{\rm
st}$ recovers the value of the half-doping, namely, $n_{\rm st} \sim 1$.
The competition of aforementioned two effects is responsible for the
temperature and hole-concentration dependences of the character of the
stripe order shown in the present study as well as
Ref. \onlinecite{yoshizawa00,katsufuji99,cheong94} through the change of
the amount of the excess carriers.

By contrast, when $n_{\rm h} =1/3$, the above two effects cooperate each
other, and $\epsilon$ is locked at 1/3 at whole $T$ range below $T_{\rm
CO}$ as shown in Fig. \ref{epsilon}.  This makes the transition of the
stripe order for $n_{\rm h} =1/3$ sharp: the intensity of the charge
order peak does not show the Debye-Waller-like behavior (not shown); the
resistivity shows a distinct step at $T_{\rm
CO}$~\cite{katsufuji99,cheong94,katsufuji96}; and a clear charge gap is
formed below $T_{\rm CO}$~\cite{katsufuji96,yamamoto98}.

% \subsubsection{Influence of the change of $n_{\rm st}$ \\
%   --- 1. influence of $n_{\rm st} \ne 1$ to charge stripe order}
%
% \subsubsection{Influence of the change of $n_{\rm st}$ \\
%   --- 2. influence of $n_{\rm st} \ne 1$ to the transport}

%\section{Summary}

%below $T_{\rm CO}$:  The variation of $n_{\rm st}$
%induces a unique temperature dependent and hole concentration dependent
%change of the character of the stripe order.

%\section*{Acknowledgement}

%This work was supported by a Grand-In-Aid for Scientific Research from
%the Ministry of Education, Science, Sports and Culture, Japan and by the
%New Energy and Industrial Technology Development Organization (NEDO) of
%Japan.

This work was supported by a Grand-In-Aid for Scientific Research from
the Ministry of Education, Culture, Sports, Science, and Technology,
Japan and by the New Energy and Industrial Technology Development
Organization (NEDO) of Japan.

\end{document}